\documentclass[12pt]{article} 
\input{epsf.tex}
\input{psfig.sty}
\newcommand{\beq}{\begin{equation}}
\newcommand{\eeq}{\end{equation}}
\newcommand{\bea}{\begin{eqnarray}}
\newcommand{\eea}{\end{eqnarray}}

\newcommand{\oh}{\frac{1}{2}}

\newcommand{\non}{\nonumber}

\newcommand{\One}{1\kern-4.5pt1}
\begin{document}

\addtolength{\baselineskip}{0.20\baselineskip}

\hfill SWAT/99/231

\hfill IFUP-TH 31/99

\hfill June 1999

\begin{center}

\vspace{48pt}

{ {\bf The Phase Diagram of the Three Dimensional Thirring Model} }

\end{center}

\vspace{18pt}

\centerline{\sl Simon Hands~$^a$
and Biagio Lucini~$^{b,c}$}

\vspace{15pt}

\centerline{$^a$Department of Physics, University of Wales Swansea,}
\centerline{Singleton Park, Swansea SA2 8PP, U.K.}
\vspace{15pt}
\centerline{$^b$Scuola Normale Superiore, Piazza dei Cavalieri 7,}
\centerline{I-56126 Pisa,  Italy.}
\vspace{15pt}
\centerline{$^c$INFN Sezione di Pisa, Via Vecchia Livornese 1291,}
\centerline{I-56010 S. Piero a Grado (Pi), Italy.}

\vspace{48pt}

\begin{center}  

{\bf Abstract}

\end{center}

We present Monte Carlo simulation results for the three dimensional Thirring
model on moderate sized lattices
using a hybrid molecular dynamics algorithm which permits an odd or non-integer 
number $N_f$ of 
fermion flavors. We find a continuous chiral symmetry breaking transition for 
$N_f\simeq3$ with critical 
exponents consistent with expectations from previous studies.
For $N_f=5$ the order of the transition is difficult to determine on the lattice
sizes explored. We present a phase diagram for the model in the 
$(1/g^2,N_f)$ plane and contrast our findings with expectations based on
approximate solutions of the continuum Schwinger-Dyson equations.



\vfill

\newpage

\section{Introduction and Algorithm}

Recent numerical investigations \cite{DH,DHM,DH2} have revealed an interesting
phase structure for the three dimensional Thirring model as a function of 
coupling $g^2$ and number of fermion species $N_f$. For sufficiently small
$N_f<N_{fc}$, there is a continuous transition at $g^2=g_c^2(N_f)$
between a weak coupling phase
in which chiral symmetry is realised in the limit $m\to0$, 
and a strong coupling one in which the
symmetry is spontaneously broken. The critical indices characterising the 
transition are distinct for $N_f=2$ and $N_f=4$, suggesting that the continuum
limits defined at the critical points define distinct interacting field
theories. For $N_f=6$, however, simulations on a $16^3$ lattice with $m=0.01$
give tentative evidence for a first order chiral transition \cite{DH2}, implying
both that no continuum limit exists in this case, and that $4<N_{fc}<6$.

Both the nature of the transition and the value of $N_{fc}$ are non-perturbative
issues, inaccessible via either a standard perturbative expansion (which is 
non-renormalisable for $d>2$), or a $1/N_f$ expansion \cite{Gomes,Hands}.
There have, however, been analytic attempts to investigate the transition via
the truncated Schwinger-Dyson (SD) equations \cite{Gomes,Itoh,HP,Sugiura}.
In the most systematic treatment \cite{Itoh}, the SD equations are solved in 
ladder approximation in the the strong coupling limit; chiral symmetry
breaking solutions are found for $N_f<N_{fc}=128/3\pi^2\simeq4.32$. Note
that in a continuum approach the value of $N_{fc}$ manifests itself as the
value of $N_f$ below which a non-trivial solution can be found, eg. by 
bifurcation theory. In a later
paper \cite{Sugiura}, Sugiura extended the solution to finite $g^2$ for
$d\in(2,4)$; for the case $d=3$ his predictions for the critical coupling as a
function of $N_f$ read:
\bea
g^2\ll1:\;\;\;g_c^2&=&{{2\pi^2}\over3}N_f\\
g^2\gg1:\;\;\;g_c^2&=&6\left[\exp\left({1\over\omega}
\left(2\pi-4\tan^{-1}\omega\right)
\right)-1\right]\label{eq:scsd}\\
&\mbox{with}&\;\;\;\;\;\;\;\omega^2(N_f)={N_{fc}\over N_f}-1\non
\eea
The essentially singular behaviour seen in
(\ref{eq:scsd}) as $N_f\nearrow N_{fc}$ is consistent with 
the existence of a conformal fixed point \cite{MirYam}. Using a different
sequence of truncations of the SD equations, however, Hong and Park \cite{HP}
found chiral symmetry breaking solutions for all $N_f$, and predicted
\begin{equation}
g_c^2\propto\exp\left({\pi^2\over16}N_f\right),
\label{eq:HP}
\end{equation}
which is manifestly non-perturbative in $1/N_f$.

The differing predictions of the SD approach motivate the use of lattice
field theory methods to address this problem. In this Letter we present
results of simulations performed with values of $N_f$ falling between
the values $N_f=2,4$ and 6 explored in previous works, in an effort both 
to map out the phase diagram in more detail, and to constrain further the value
of $N_{fc}$. To start with, let us define the lattice action: 
\bea
        S &=& \oh \sum_{x\mu i} \bar\chi_i(x) \eta_\mu(x)
        \biggl[(1+iA_\mu(x)) \chi_i(x+\hat\mu)
        -(1-iA_\mu(x-\hat\mu))\chi_i(x-\hat\mu)\biggr] \non \\
          & & + m \sum_{xi} \bar\chi_i(x) \chi_i(x) +
        \frac{N}{4g^2} \sum_{x\mu} A_\mu^2(x)\label{eq:Slat}\\
          &\equiv& \sum_{xyi}\bar\chi_i(x){\cal M}_{ij}[A](x,y)\chi_j(y)
          +{N\over{4g^2}}\sum_{x\mu}A_\mu^2(x),\non
\eea
where the indices $i,j$ run over $N$ flavors of staggered lattice fermion.
The $A_\mu$ are real auxiliary vector fields defined on the lattice links;
gaussian integration over $A_\mu$ yields a form of the action with explicit
four-fermion couplings \cite{DHM,DH2} (other variants of the lattice-regularised
Thirring model are discussed in \cite{DHM}). Analysis of the spin-flavor content
of staggered lattice fermions \cite{BB} reveals that in 3 dimensions the number
of continuum four-component physical flavors is given by $N_f=2N$. Therefore
use of the above action limits us to even $N_f$. If, however, the fermions
are integrated out, then the resulting effective action for the $A_\mu$
fields reads:
\begin{equation}
S_{eff}={N_f\over{8g^2}}\sum_{x\mu}A_\mu^2(x)-{N_f\over2}\ln\mbox{det}{\cal
M}[A].
\end{equation}

\begin{figure}[htb]
\psdraft
\centerline{
\setlength\epsfxsize{300pt}
\epsfbox{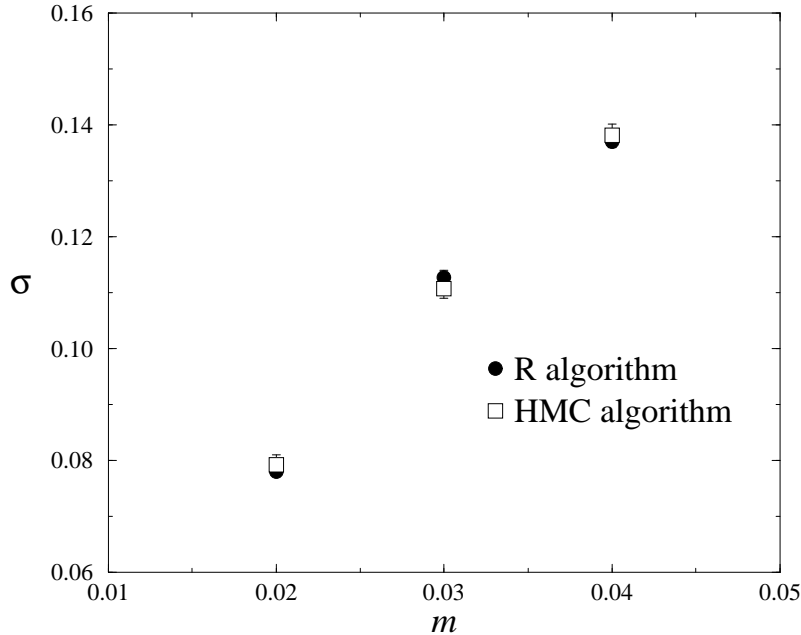
}}
\psfull
\caption{Chiral condensate (here denoted $\sigma$) vs. $m$ on a $12^3$ lattice
for $N_f=4$ with $1/g^2=1.0$.
\label{fig:compare}}
\end{figure}
The effective action, though non-local, has an analytic dependence on $N_f$, 
and hence can be employed for odd or non-integer 
values.\footnote{Note that the form of (\ref{eq:Slat}) permits even-odd
partitioning, which means that we can simulate integer powers of
$\mbox{det}{\cal M}$,
rather than $\mbox{det}{\cal M^{\dagger}M}$, with a local action.} It can be
simulated using a hybrid molecular dynamics algorithm, which evolves the
$\{A\}$ configuration through a fictitious time $\tau$ by a combination of 
microcanonical and Langevin dynamics. In the limit of timestep $\Delta\tau\to0$
the $\{A\}$ are distributed according to the equilibrium ensemble.
In this work we have implemented the R algorithm of Gottlieb {\it et al\/}
\cite{Gottlieb}; in this case the systematic errors are 
$O(N^2\Delta\tau^2)$ \cite{Gottlieb,HK}. We found that on a $12^3$ lattice 
a value $\Delta\tau=0.01$ was sufficient, with a mean interval
$\bar\tau=1.0$ between refreshments. Our results typically arise from
averages over 500 units of $\tau$. 
Fig.~\ref{fig:compare} shows a 
comparison between chiral condensate $\langle\bar\chi\chi\rangle$ data
obtained for $N_f=4$, for various values
of the bare mass $m$, using both the hybrid algorithm and a hybrid Monte 
Carlo algorithm \cite{DHM}, which in principle is free from systematic error.

In Sec. \ref{sec:fixedg} we will present results from simulations with 
variable $N_f$ at a fixed
coupling $1/g^2=1.0$. The phase transition is located and found to fall at 
a value of $N_f$ intermediate to the cases studied in earlier work
\cite{DHM,DH2}. Fits to a power law equation of state ansatz yield critical
exponents consistent with this picture. In Sec. \ref{sec:fixedN} we present
results obtained with variable $g^2$ at a fixed $N_f=5$, the goal being to
determine if the transition remains continuous, or whether there is evidence
for coexisting phases signalling a first order transition, as observed for
$N_f=6$ \cite{DH2}. Our conclusions, and the resulting phase diagram, are given
in Sec. \ref{sec:conc}.

\section{Numerical Simulations} 
\subsection{Fixed $1/g^2=1.0$}\label{sec:fixedg}

\begin{figure}[htb]
\psdraft
\centerline{
\setlength\epsfxsize{300pt}
\epsfbox{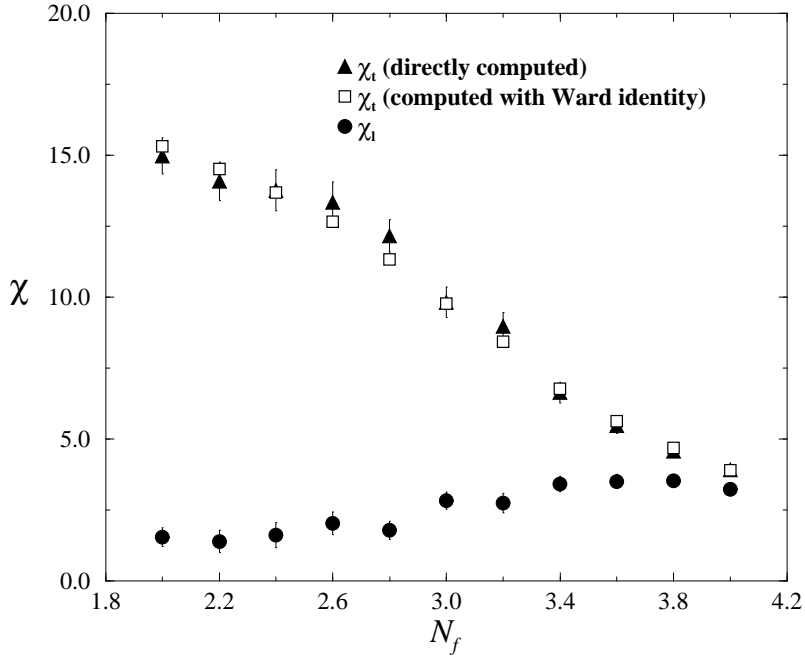
}}
\psfull
\caption{Transverse and longitudinal susceptibilities 
 vs. $N_f$ on a $12^3$ lattice
with $1/g^2=1.0$, $m=0.02$.
\label{fig:suscept}}
\end{figure}
\begin{figure}[tb]
\psdraft
\centerline{
\setlength\epsfxsize{300pt}
\epsfbox{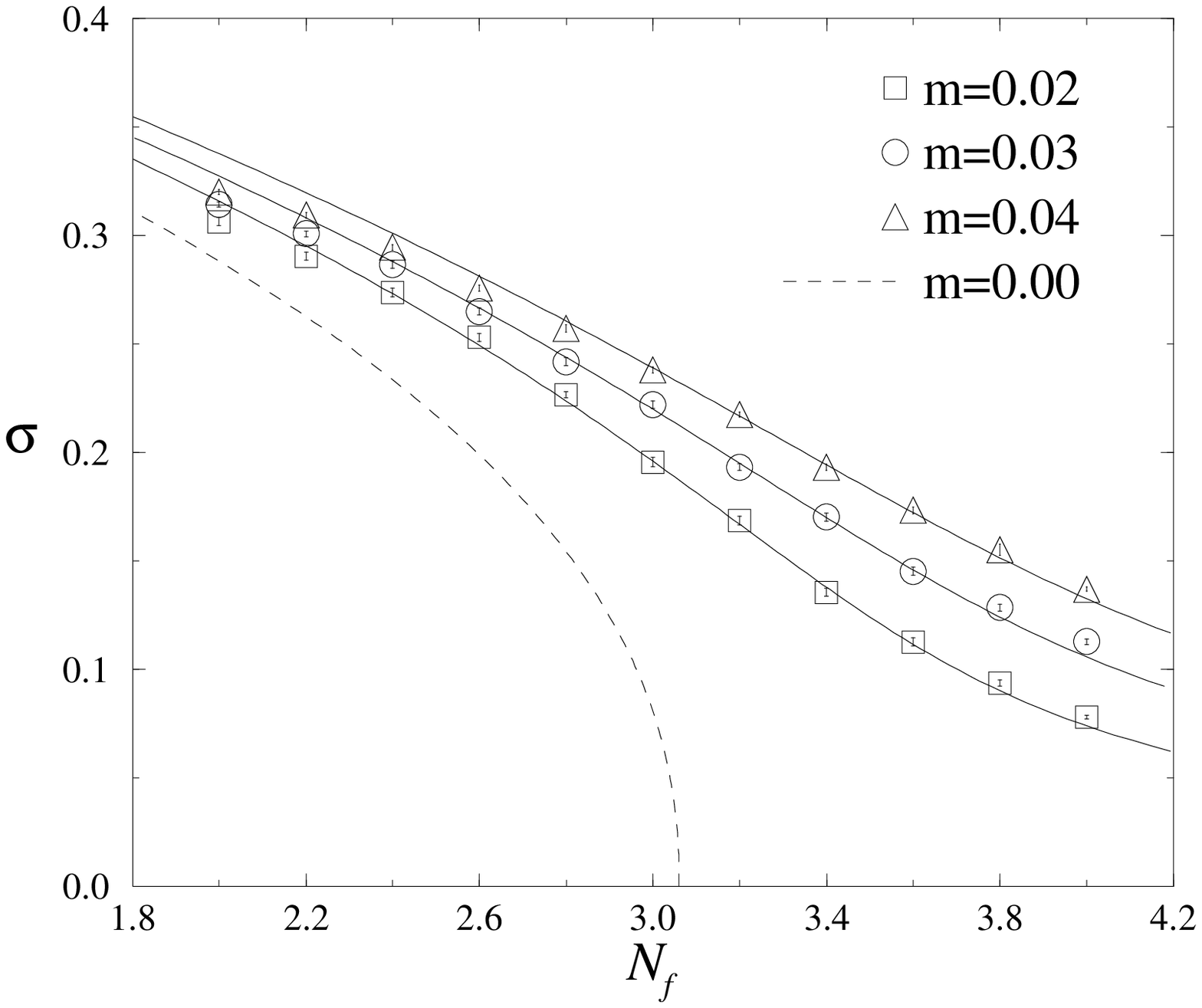
}}
\psfull
\caption{Chiral condensate $\sigma$  vs. $N_f$ on a $12^3$ lattice
with $1/g^2=1.0$, showing fits to the equation of state (\ref{eq:eos}).
The dashed line shows the fit in the chiral limit.
\label{fig:condfit}}
\end{figure}
Using the algorithm as described above we performed simulations with
variable $N_f$ on a $12^3$
lattice at fixed $1/g^2=1.0$, with bare mass $m=0.04,0.03,0.02$.
In Fig.~\ref{fig:suscept} we plot the transverse and longitudinal
susceptibilities $\chi_{t,l}$, using the definitions given in \cite{DHM,DH2}.
As described there, $\chi_t$ can be calculated either
from the integrated two-point pion propagator, or from 
$\langle\bar\chi\chi\rangle$ via the axial Ward identity: the two methods give
compatible results, the latter being the less noisy.
These quantities are related to the inverse square masses of
respectively the pseudoscalar and scalar bound states. For $N_f\geq3.6$ they
are roughly equal, indicating that the scalar and pion states are degenerate,
and chiral symmetry is realised. For $N_f\leq2.8$, in contrast,
$\chi_t\gg\chi_l$, indicating that the pion has become light as expected for
a pseudo-Goldstone mode in a phase of broken chiral symmetry.
Fig.~\ref{fig:suscept} thus offers evidence for a chiral symmetry breaking phase
transition in the region $N_f\simeq3.0$.

Fig.~\ref{fig:condfit} shows results for the order parameter
$\langle\bar\chi\chi\rangle$ as a function of $N_f$, together with fits to 
a renormalisation-group inspired equation of state \cite{DHM} of the form
\begin{equation}
m=At\langle\bar\chi\chi\rangle^{\delta-1/\beta}+B\langle\bar\chi\chi\rangle^
\delta,\label{eq:eos}
\end{equation}
where $\delta,\beta$ are conventional critical exponents, and the parameter
$t$ expresses the distance from criticality. In previous work at fixed $N_f$,
variable $1/g^2$ we have defined $t=1/g^2-1/g_c^2$; here we define
\begin{equation}
t\equiv N_f-N_{f*},
\end{equation}
where $N_{f*}$  denotes the value of $N_f$ at the transition.
Experience from previous studies \cite{DH,DHM,DH2} has shown 
that equations of state of the sort (\ref{eq:eos}) provide adequate
descriptions of the data in the vicinity of the transition, and that the 
resulting fitted exponents do not differ much from those extracted by more
sophisticated finite volume scaling analyses.
Also following previous work we fix $\beta$ by the requirement
$\delta-1/\beta=1$, in order to stabilise the fitting procedure; we refer to 
the resulting four parameter fit as fit I. 
The results obtained by fitting data for $N_f\in[3.0,3.6]$ are given in
Table~\ref{tab:fit}. Also shown is a five parameter fit II over the 
same range for which the 
constraint on $\beta$ was relaxed.
\begin{table}[ht]
\setlength{\tabcolsep}{1.5pc}
\caption{Results for $1/g^2=1.0$ from fits to (\ref{eq:eos}) on a $12^3$ 
lattice.}
\label{tab:fit}
\begin{tabular*}{\textwidth}{@{}l@{\extracolsep{\fill}}llll}
\hline
        & Parameter      &   Fit I     &   Fit II     \\
\hline
        & $N_{f*}$       &  3.060(30)  &  3.060(44)     \\
        & $A$            &  0.264(7)   &  0.253(8)     \\
        & $B$            &  3.9(9)     &  4.4(1.3)    \\
        & $\delta$       &  3.14(17)   &  3.22(24)    \\
        & $\beta$        &  ---        &  0.44(5)    \\
        & $\chi^2$/d.o.f. &  1.3        &  1.0         \\
\hline
\end{tabular*}
\end{table}
A number of comments are in order. First, the constraint of fit I appears
to be well satisfied by the data, since from fit II the fitted value 
of $\delta-1/\beta=0.97(8)$. Second, the fitted values of the exponents 
fall between those found for $N_f=2$ and $N_f=4$ from runs with 
variable $1/g^2$ \cite{DH}, being numerically closer to the $N_f=4$ case: 
the results for these from comparable fits are summarised 
in Table~\ref{tab:N24}.
This supports the idea that $g_c^2(N_f)$ 
is a critical line of fixed points along which critical exponents vary smoothly,
with the numerical value of the
exponents independent of the direction from within the $(g^2,N_f)$ plane from
which the critical line is approached. Third, as a note of caution, it should be
noted that the criterion for the range of $N_f$ over which the fit 
was made was purely on the basis on minimising the $\chi^2$ per degree of
freedom, as in \cite{DH,DHM,DH2}. Inspection of Fig. \ref{fig:condfit} 
suggests that the resulting fits may not describe the broken phase
particularly well, and indeed fits which included smaller $N_f$ values 
resulted in systematically larger values of $\delta$. Our philosophy is 
to regard (\ref{eq:eos}) merely as a phenomenological decription of the data;
results from comparable fitting procedures applied to different models can thus
be used to establish trends regardless of whether the 
true equation of state is ultimately  of the form (\ref{eq:eos}) or not.
\begin{table}[ht]
\setlength{\tabcolsep}{1.5pc}
\caption{Comparison with critical parameters for the $N_f=2$ and 4 models from
a $12^3$ lattice from ref. [1].}
\label{tab:N24}
\begin{tabular*}{\textwidth}{@{}l@{\extracolsep{\fill}}llll}
\hline
        & Parameter      &   $N_f=2$   & ``$N_f=3.06$''  &   $N_f=4$     \\
\hline
        & $1/g_c^2$ (fit I) &  1.94(4)  &    1.00(1)     &  0.66(1)     \\
        & $\delta $ (fit I) &  2.68(16) &    3.14(17) &   3.43(19)    \\
        & $\beta$  (fit II)  &  0.71(9)  &   0.44(5)  &    0.38(4)    \\
\hline
\end{tabular*}
\end{table}

\subsection{Fixed $N_f=5$}\label{sec:fixedN}

\begin{figure}[htb]
\psdraft
\centerline{
\setlength\epsfxsize{300pt}
\epsfbox{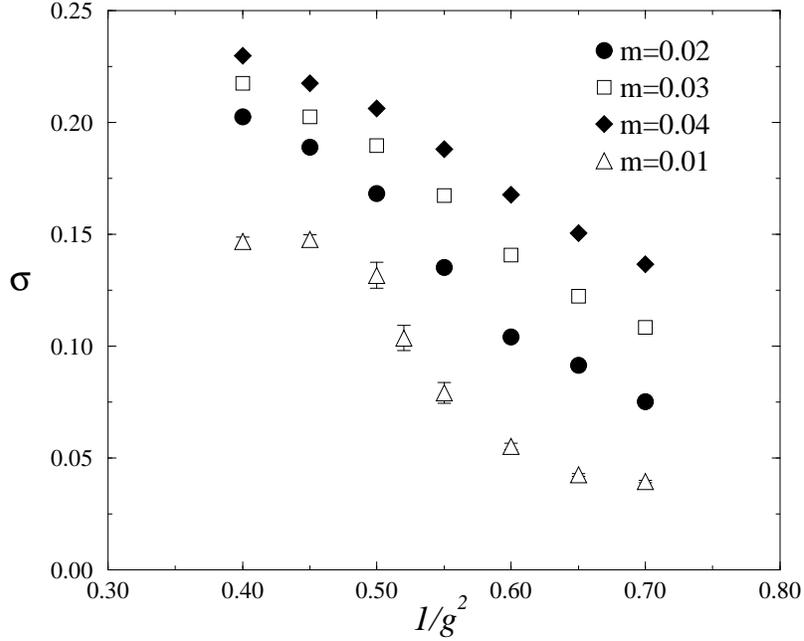
}}
\psfull
\caption{Chiral condensate $\sigma$ as a function of $1/g^2$ for
$N_f=5$ on a $12^3$ lattice.
\label{fig:condnf5}}
\end{figure}
With the confidence that simulations with a non-integer number of staggered
flavors yield results consistent with our expectations for $N_f\simeq3$, we
then conducted a series of simulations on $12^3$ lattices with similar
parameters but this time with $N_f=5$, a value which previous
studies suggest is close to $N_{fc}$. 
Our results for
$\langle\bar\chi\chi\rangle$ are shown in Fig.
\ref{fig:condnf5}. We have included results from simulations with $m=0.01$,
although it is likely that these are badly finite-volume affected; however,
they do reveal evidence of discontinuous behaviour for $1/g^2\simeq0.5$.
Fits to the equation of state (\ref{eq:eos}), with the parameter $t$ now given 
by $1/g^2-1/g_c^2$, were attempted, but were less successful than
those of the previous section, in general either resulting in an
unacceptably large $\chi^2$ or not reaching convergence. In any case, the
resulting fitted parameters are strongly dependent on the fit range, and
inclusion of the $m=0.01$ points makes  fit quality much worse.
Our best results, coming from fitting
data with $m\in[0.04,0.02]$, $1/g^2\in[0.4,0.6]$, are shown in Tab.
\ref{tab:fit5}.
\begin{table}[ht]
\setlength{\tabcolsep}{1.5pc}
\caption{Results for $N_f=5$ from fits to (\ref{eq:eos}) on a $12^3$ 
lattice.}
\label{tab:fit5}
\begin{tabular*}{\textwidth}{@{}l@{\extracolsep{\fill}}llll}
\hline
        & Parameter      &   Fit I     &   Fit II     \\
\hline
        & $1/g_c^2$      &  0.437(4)   &  0.47(2)     \\
        & $A$            &  1.15(3)    &  0.5(2)     \\
        & $B$            &  190(40)    &  22(20)    \\
        & $\delta$       &  5.6(1)     &  4.1(6)    \\
        & $\beta$        &  ---        &  0.29(5)    \\
        & $\chi^2$/d.o.f. &  2.7        &  2.3         \\
\hline
\end{tabular*}
\end{table}
The fitted values of $\delta$ and $\beta$ are consistent with the trends of
Tab. \ref{tab:N24}, although it 
is interesting to note that the fitted
value of $\delta-1/\beta=0.6(2)$, indicating that in contrast
to previous fits the data do not 
satisfy the usual constraint in this case. 
It could be argued that this disfavours the fit, since 
the combined assumptions of the degeneracy of scalar
and pseudoscalar bound states in the chiral limit of the symmetric phase, plus
the applicability of a power-law equation of state (\ref{eq:eos}), naturally
predict $\delta-1/\beta=1$ \cite{DHM,KKW+}.

\begin{figure}[htb]
\psdraft
\centerline{
\setlength\epsfxsize{300pt}
\epsfbox{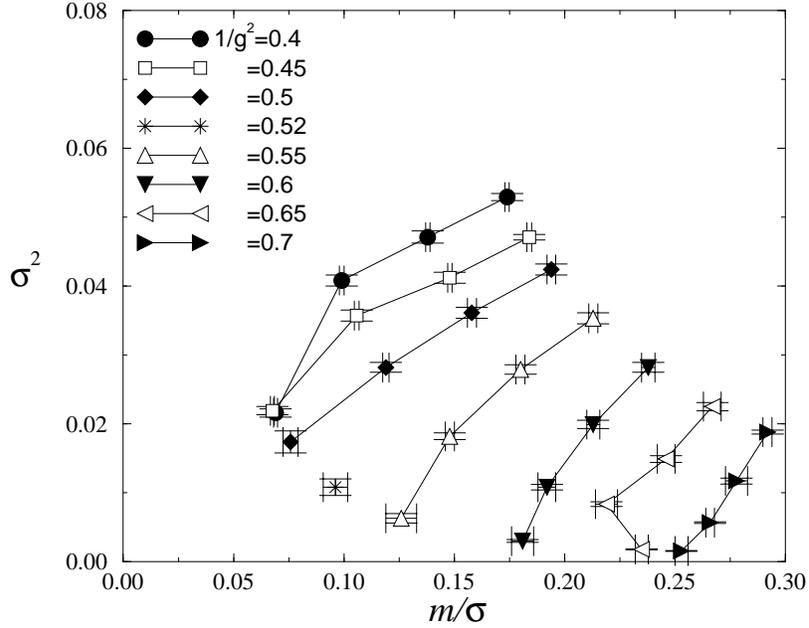
}}
\psfull
\caption{Fisher plot of $\sigma^2$ vs. $m/\sigma$
for $N_f=5$ on a $12^3$ lattice.
\label{fig:fishnf5}}
\end{figure}
\begin{figure}[htb]
\psdraft
\centerline{
\setlength\epsfxsize{300pt}
\epsfbox{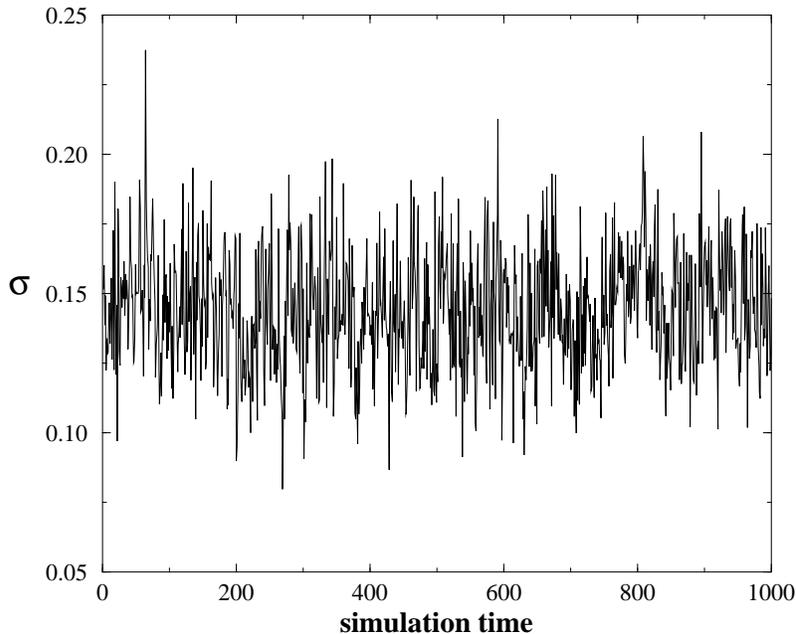}}
\psfull
\caption{Time history of $\sigma$ for
$N_f=5$, $m=0.01$ on a $16^3$ lattice.
\label{fig:t47}}
\end{figure}

To gain more insight we present the same data in the form of a Fisher plot
(ie. $\langle\bar\chi\chi\rangle^2$ vs. $m/\langle\bar\chi\chi\rangle$) 
in Fig. \ref{fig:fishnf5}. This plot is devised to yield trajectories
of constant
$1/g^2$ which intersect the vertical axis in the chiral limit in the broken
phase, and the horizontal axis in the symmetric phase. Departures from 
the mean field indices $\delta=3,\beta={1\over2}$ are revealed as curvature
in the lines, and departures from $\delta-1/\beta=1$ are revealed by variations
in the sign of the curvature between the two phases. If we ignore the $m=0.01$
points, which are those closest to the horizontal axis, then the plot suggests
a critical $1/g_c^2\simeq0.5$, with tentative evidence for $\delta-1/\beta<1$.
Inclusion of the low mass points, however, suggests an accumulation of the
constant coupling trajectories
around a line which if continued would intercept the
horizontal axis. This is similar to the Fisher plot for $12^3$
data for the model with $N_f=6$, 
shown in fig. 6 of \cite{DHM}, and suggests similarities between the two cases. 
For $N_f=6$, tentative evidence
for tunnelling between coexisting vacua on simulations
with $m=0.01$ on a
$16^3$ lattice in the critical region
was presented in \cite{DH2}, consistent with the chiral
transition being first order. We tested this possibility by performing 
long $N_f=5$ simulations on a $16^3$ lattice for $m=0.01$ and
$1/g^2=0.45$, 0.47 (these runs requiring $\Delta\tau=0.002$):
a time history for the latter is shown in Fig. \ref{fig:t47}.
Whilst the fluctuations are large, with excursions of $O(50\%)$
about the mean, there is no convincing evidence for a two-state signal.
Double gaussian fits to the histograms of the binned data, which revealed
coexisting states via twin peaks in \cite{DH2}, were this time 
statistically indistinguishable from simple single gaussians.

In summary, whilst there seems to be clear evidence for a chiral transition
with $1/g^2=0.50(5)$, the simulations performed to date are unable to 
determine the order.
Fits based on the assumption of a continuous transition, whilst falling into the
broad trend of existing data, are unsatisfactory and fail to satisfy the
constraint $\delta-1/\beta=1$. A search for evidence of metastability
consistent with a first order transition, on the other hand, proved negative.

\section{The Phase Diagram}\label{sec:conc}

\begin{figure}[htb]
\psdraft
\centerline{
\setlength\epsfxsize{300pt}
\epsfbox{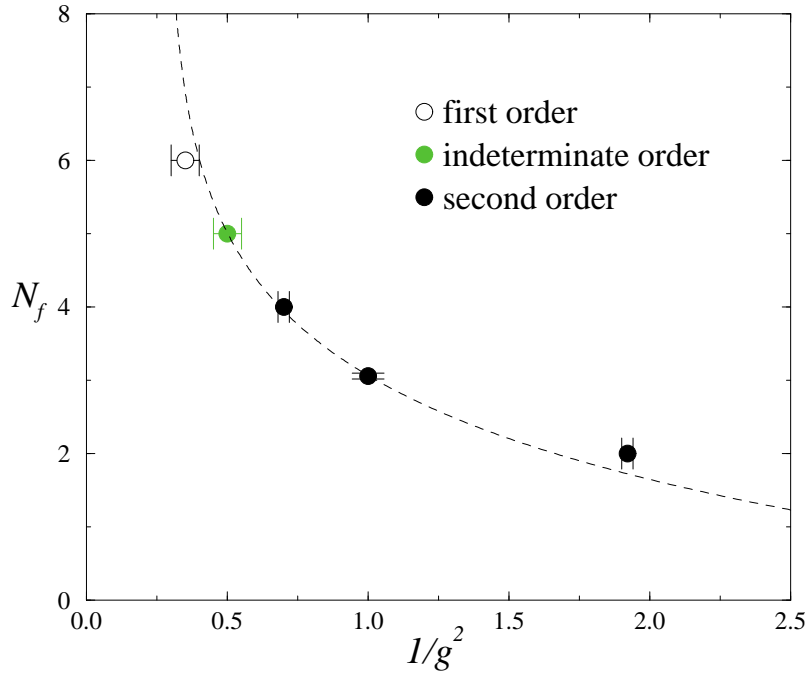
}}
\psfull
\caption{Phase diagram for the three dimensional Thirring model;
the chirally broken phase is to the lower left. The dashed line shows 
a fit to (\ref{eq:HPf}).
\label{fig:phased}}
\end{figure}

In this Letter we have used a fresh simulation algorithm to explore
non-integer numbers of staggered fermion flavors, and found results broadly
consistent with earlier studies, namely that for $N_f\simeq3$ the transition
is continuous with critical exponents intermediate between those of
$N_f=2$ and $N_f=4$ models, and that for $N_f=5$ the order of the transition is
difficult to determine, but shares features in common with the $N_f=6$
transition, believed to be first order. This change in order of the 
transition with increasing 
$N_f$ has also been observed in studies of $\mbox{QED}_4$
\cite{QED4N}.
The results also support the view that the partition function can be regarded
as analytic in $N_f$, and that the non-locality of the fermionic action
for $N_f$ odd has no severe consequences.
In this final section we collect together these new results 
with the best estimates obtained from previous
finite volume scaling studies
in \cite{DHM,DH2}, to plot the phase diagram of the Thirring model on the 
$(1/g^2,N_f)$ plane in the chiral limit. The result is shown in Fig.
\ref{fig:phased}. The shading of the points indicates the suspected order
of the transition. The picture we have established is consistent with 
a critical line of fixed points, 
along which critical exponents vary smoothly, up to some 
$N_{fc}\simeq5$, whereupon the transition becomes first order.

It is work comparing our findings to the analytic approximations of 
Refs. \cite{HP} and \cite{Sugiura}. Certainly the curvature of our 
critical line $1/g_c^2(N_f)$ is consistent with both predictions
(\ref{eq:scsd}) and (\ref{eq:HP}). A detailed comparison with the former,
which predicts a conformal fixed point at 
$N_f=N_{fc}\simeq4.32$ 
is hampered by the difficulties in identifying the strong coupling
limit in the lattice model, since there is an additive renormalisation relating 
$1/g^2_{CONT}$ to $1/g^2_{LATT}$ due to non-conservation of the interaction
current in the lattice model \cite{DH,DHM}. Therefore we are forced to 
identify $N_{fc}$ with that value at which the transition becomes first order,
at which point no continuum limit exists for the lattice model. Our estimate 
for $N_{fc}$ is thus roughly consistent; however,
the trends in the exponents extracted from the lattice studies,
summarised in Table. \ref{tab:crit} do not
support a conformal fixed point. At such a point we expect the exponent 
$\delta$ to take the value 1, as for an asymptotically-free theory. 
The exponent $\eta$ is related to the anomalous dimension of the
composite operator $\bar\chi\chi$ by $\eta=d-2\gamma_{\bar\chi\chi}$
\cite{HKM}, which in 
turn is predicted at the fixed point to be given by \cite{Sugiura}:
\begin{equation}
\gamma_{\bar\chi\chi}={{d-2}\over2}.
\end{equation}
At a conformal fixed point 
$\eta$ should thus take the value 2. 
\begin{table}[ht]
\setlength{\tabcolsep}{1.5pc}
\caption{Critical exponents as a function of $N_f$.}
\label{tab:crit}
\begin{tabular*}{\textwidth}{@{}l@{\extracolsep{\fill}}llll}
\hline
        & $N_f$      &   $\delta$  &   $\eta$     \\
\hline
        &  2      &  2.75(9)   &  0.60(2)     \\
        & 3.06(3)   &  3.14(17)    &  ---    \\
        &  4  &  3.76(14)    &  0.26(4)   \\
        &  5  &  5.6(1)?    &  ---   \\

\hline
\end{tabular*}
\end{table}
Finally, it is interesting to note that a fit of the form
\begin{equation}
1/g_c^2=A+B\exp\left(-{\pi^2\over16}N_f\right),
\label{eq:HPf}
\end{equation}
motivated by Eq. (\ref{eq:HP}) adjusted to allow for the renormalisation of
$1/g^2$,
can be made to the data shown in Fig. \ref{fig:phased}; the fit is particularly
successful if the $N_f=2$ point is excluded, in which case the resulting
coefficients are $A=0.28(2)$, $B=4.7(2)$, with $\chi^2$/d.o.f.=0.6. 
This suggests
that the scenario of \cite{HP}, in which chiral symmetry breaking is predicted
for all $N_f$ and hence that $N_{fc}=\infty$, cannot be excluded by the lattice
studies to date.

\section*{Acknowledgements}

This project was supported in part 
by the TMR-network ``Finite temperature phase transitions in particle
physics'' EU-contract ERBFMRX-CT97-0122. We are greatly indebted to the
CRT Computer Center of ENEL (PISA) for collaboration in the use of their
CRAY YMP-2E. One of us (B.L.) would like to thank the Physics Department
of Swansea for kindly hospitality.

\clearpage

\end{document}